\PassOptionsToPackage{svgnames}{xcolor}
\documentclass[sigconf]{acmart}
\usepackage[utf8]{inputenc}
\usepackage{xspace}
\usepackage{enumitem}
\usepackage{booktabs}
\usepackage{multirow}
\usepackage{mathptmx}                 % use matching math font
\usepackage{graphicx}
\usepackage{bm}
\DeclareUnicodeCharacter{FF0C}{,}

\newcommand{\mymethod}{CHORDination\xspace}
\newcommand{\etal}{et~al.\xspace}

\newcommand{\revised}[1]{\textcolor{black}{#1}}
\newcommand{\hide}[1]{}
\AtBeginDocument{%
  }

\setcopyright{acmlicensed}
\copyrightyear{2024}
\acmYear{2024}
\acmDOI{XXXXXXX.XXXXXXX}

\acmConference[Conference acronym 'XX]{Make sure to enter the correct
  conference title from your rights confirmation emai}{June 03--05,
  2024}{Hsinchu, Taiwan}
\acmISBN{978-1-4503-XXXX-X/18/06}

\begin{document}

\title{\mymethod: Evaluating Visual Design Choices in Chord Diagrams for Network Data}

\author{Kai Wang}
\authornote{The first two authors contributed equally to this research.}

\affiliation{%
  \institution{Xi'an Jiaotong-Liverpool University}
  \city{Suzhou}
  \country{China}
}
\orcid{0009-0008-0982-8919}

\author{Shuqi He}
\authornotemark[1]
\affiliation{%
  \institution{Xi'an Jiaotong-Liverpool University}
  \city{Suzhou}
  \country{China}
}
\orcid{0009-0002-6365-8806}

\author{Wenlu Wang}
\affiliation{%
  \institution{Xi'an Jiaotong-Liverpool University}
  \city{Suzhou}
  \country{China}
}
\orcid{0009-0005-5548-5732}

\author{Jinbei Yu}
\affiliation{%
  \institution{Xi'an Jiaotong-Liverpool University}
  \city{Suzhou}
  \country{China}
}

\author{Yu Liu}
\affiliation{%
  \institution{Xi'an Jiaotong-Liverpool University}
  \city{Suzhou}
  % \state{Jiangsu}
  \country{China}
}

\author{Lingyun Yu}
\authornote{Corresponding author}
\affiliation{%
  \institution{Xi'an Jiaotong-Liverpool University}
  \city{Suzhou}
  % \state{Jiangsu}
  \country{China}
}
\orcid{0000-0002-3152-2587}


\begin{abstract}
Chord diagrams are widely used for visualizing data connectivity and flow between nodes in a network. They are effective for representing complex structures through an intuitive and visually appealing circular layout. While previous work has focused on improving aesthetics and interactivity, the influence of fundamental design elements on user perception and information retrieval remains under-explored. In this study, we explored the three primary components of chord diagram anatomy, namely the nodes, circular outline, and arc connections, in three sequential experiment phases. In phase one, we conducted a controlled experiment (N=90) to find the perceptually and information optimized node widths (narrow, medium, wide) and quantities (low, medium, high). This optimal set of node width and quantity sets the foundation for subsequent evaluations and were kept fixed for consistency. In phase two of the study, we conducted an expert design review for identifying the optimal radial tick marks and color gradients. Then in phase three, we evaluated the perceptual and information retrieval performance of the design choices in a controlled experiment (N=24) by comparing four chord diagram designs (baseline, radial tick marks, arc color gradients, both tick marks and color gradients).
Results indicated that node width and quantity significantly affected users' information retrieval performance and subjective ratings, whereas the presence of tick marks predominantly influenced subjective experiences.
Based on these findings, we discuss the design implications of these visual elements and offer guidance and recommendations for optimizing chord diagram designs in network visualization tasks. 
\end{abstract}






\maketitle

\section{Introduction}
Network data is ubiquitous across many research domains, including social networks \cite{D3Migration,Nicholas2014InteractiveVO,Abel2014QuantifyingGI}, biological systems \cite{Finnegan2019UsingAC}, and transportation systems \cite{Domains,6876029}. These complex network scenarios require effective visualization techniques to represent the interconnected relationships and flows. Among various techniques, chord diagrams stand out for their intuitive circular layout and ability to display bidirectional relationships compactly \cite{FusionCharts}.

\begin{figure}[htbp]
    \centering
    \includegraphics[width=0.8\columnwidth]{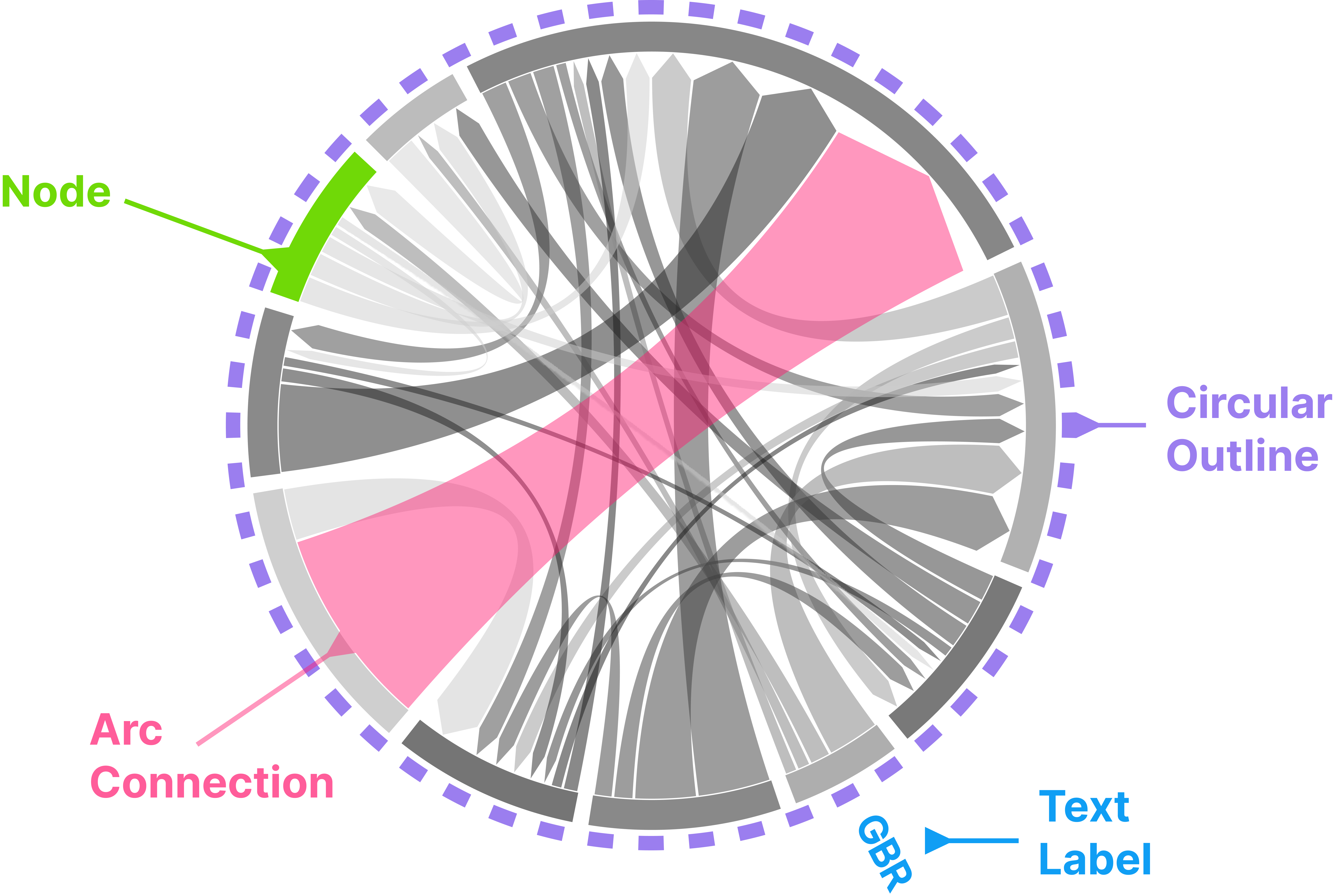}
    \caption{The anatomy of a chord diagram, with key elements highlighted and labeled.}
    \Description{A chord diagram illustrating the positions of key elements: `Node' in green, which are segments located along the circular outline, `Arc Connection' in pink spanning the central area, `Circular Outline' in purple encircling the entire diagram, and `Text Label' in blue positioned around the outer edge. The diagram demonstrates the spatial arrangement of these elements within the chord structure.}
    \label{fig:chordanatomy}
\end{figure}

The anatomy of a chord diagram consists of several key components (\autoref{fig:chordanatomy}). The \textit{circular outline} forms the backbone of a chord diagram, providing a structural foundation. Data entities are represented as segments called \textit{nodes} along this outline. Chords, or \textit{arc connections}, are the links that connect between nodes. These elements, combined with color and text labels, create visually appealing and space-efficient designs for network visualization.

Despite their popularity, chord diagrams have some known drawbacks such as visual clutter and difficulty in accurately perceiving connection weights due to overlapping connections. Gutwin et al.'s work \cite{Gutwin:2023:SFC} suggested that chord diagrams are less effective than its alternative sankey diagrams. However, how the design variations of chord diagrams influence user perception and information seeking performance remains under-explored. Most prior research has focused on improving the aesthetic and interactive features of these diagrams rather than systematically evaluating the impact of basic design elements \cite{Nicholas2014InteractiveVO}. For example, Haghnazar \etal \cite{Koochaksaraei2017ANV}, Kakaraparty \cite{color} and Kriebel \cite{Flourish} employed ribbon design and color design for aesthetic improvements. Finnegan \etal \cite{Finnegan2019UsingAC} and Kriebel \cite{Flourish} implemented hover highlighting for interactivity.

To address this research gap, we chose to examine one primary design consideration per component: the \textbf{node width and quantity} of the node segments, the use of \textbf{color gradients} on the arc connections, and the presence of \textbf{radial tick marks} on the circular outline. Specifically, we formulated the following research questions:

\begin{itemize}
    \item \textbf{RQ1}: How do variations in node width and quantity affect the readability of chord diagrams?
    \item \textbf{RQ2}: What are the key design considerations for color gradients and tick marks when added in chord diagrams?
    \item \textbf{RQ3}: How do color gradients and radial tick marks influence the perception of chord diagrams?
\end{itemize}

We designed three experimental phases to investigate these factors and derive optimal design choices. The first phase focused on determining the optimal node parameters for presentation and foundational readability (N=$90$). The second phase aimed to narrow down the tick marks and color gradients design with expert review. The third phase examined the effects of the pre-selected radial tick marks and color gradients designs (N=$24$). Ultimately, \mymethod, aims to derive practical design guidelines by systematically investigating how design components influence user perception and information acquisition in chord diagrams.

\section{Related Work}
\subsection{Network Data Visualization}
As an important branch of information visualization, network data visualization shows data relationships without lengthy explanations. Techniques include matrix charts, node-link diagrams, word clouds, and alluvial diagrams \cite{4rela}. Network data is typically represented by nodes and edges, where nodes represent entities and edges represent relationships between these entities \cite{5rela}. 

Node-link diagrams use geometric shapes to represent nodes and lines to represent edges \cite{7rela}. For instance, edges are typically drawn as straight lines in social network visualizations, whereas more complex geometries and curved edges are used in dense networks to reduce occlusion \cite{8rela}. However, node-link diagrams are criticized for producing visual clutter with complex data \cite{9rela}.

Adjacency matrix representations are particularly suitable for identifying clusters and communities within networks \cite{7rela, 10rela}. However, as matrices do not directly visualize paths or connections, they may be less straightforward than node-link diagrams \cite{11rela}.

Node-link diagrams and matrices can be combined together, allowing for the display of both the global structure and local details of networks \cite{7rela}. This method is advantageous in handling complexity in the inter-community structures and local densities \cite{12rela}. However, implementing it requires more sophisticated algorithms and computational resources \cite{13rela}.

\subsection{Chord Diagrams for Network Data}
Chord diagrams originated as a variant of Cartesian graphs, known as radial diagrams \cite{relatedwork_8_burch2013benefits}. Its circular appearance offers better scalability \cite{relatedwork_9_islam2021circle} to accommodate more data within the same space. Additionally, due to the centralized nature of circular diagrams, users' visual attention tends to focus on the center of the circle \cite{relatedwork_10_caistudy,relatedwork_11_he2024visual}. In a chord diagram, links can be either bidirectional or unidirectional. For unidirectional links, the direction of the chord represents the flow of data, while bidirectional links are more complex. This experiment adopts unidirectional links for simplicity. 
%In terms of directionality, the color of the connection remains consistent with the originating node \cite{relatedwork_7_lee2023using}. 
%The width of the bidirectional connection on both ends of chord is different \cite{relatedwork_2_iturbe2016visualizing,relatedwork_3_jalali2016supporting,relatedwork_5_cheng2023global}, while unidirectional connections has a consistent width at both ends \cite{Gutwin:2023:SFC,relatedwork_7_lee2023using}.

\subsection{Visual Design Choices}
Node quantity is a critical design consideration for network visualizations. In a comprehensive evaluation by Komarek et al. \cite{relatedwork_4_komarek2015network}, it was tested that chord diagrams can display up to 100 sets of data while maintaining aesthetics and readability. 

Moreover, different aspects of color choices can enhance chord diagram readability. Specific color choices can be reserved for encoding anomalies or specific features \cite{relatedwork_2_iturbe2016visualizing}. Additionally, color settings such as transparency \cite{relatedwork_1_10.1145/2554850.2554886} and brightness \cite{relatedwork_7_lee2023using} can be adjusted to reveal overlapping elements and highlight differences.
% For instance, Iturbe et al. \cite{relatedwork_2_iturbe2016visualizing} used chords with specific color representations for indicating anomalies in industrial networks. 
% In addition, the transparency can be adjusted to reveal overlapping elements and enhance visual support \cite{relatedwork_1_10.1145/2554850.2554886}. The brightness of colors can also be modified. When building bibliometric analyses using chord diagrams \cite{relatedwork_7_lee2023using}, different references are matched with chord color hues, while variations in brightness are used to represent different countries in a straightforward manner.

More recently, tick marks have been studied in visualizations for providing a visual reference and enhancing accuracy estimation. For example, Kosslyn et al. \cite{relatedwork_12_kosslyn2006graph} suggested that adding dense tick marks can calibrate axis. Likewise, it has been demonstrated in donut charts that tick marks can improve accuracy of estimating proportions \cite{relatedwork_10_caistudy}. Teng-Yun Ch et al. \cite{relatedwork_5_cheng2023global} used outer ring tick marks for facilitating value comparisons in chord diagrams. 

\subsection{Relevant Tasks \& Performance Metrics}
Chord diagrams are evaluated through tasks related to visual search, numerical relationship comparison, and path following. Studies suggested that strong color contrasts and consistent layouts avoiding acute-angle crossings help users locate information quickly and improves readability \cite{02rela, 03rela}. Visual comparison is crucial in data analysis as users need to draw meaningful conclusions from comparing quantities \cite{04rela}. Clear and easily comparable graphic designs such as aligned bar charts or proportionate pie charts can help users perform quantity comparisons more accurately \cite{05rela}. Path following in flow diagrams involves tracing links between entities. Using straight lines instead of curves and reducing line crossings can significantly improve path-following accuracy \cite{06rela, 07rela}.
% Studies also showed that path segments on a smooth curve are easier to see, and the degree of ``bendiness'' correlates with the time required to interpret the path \cite{03rela}.

User performance metrics are essential for assessing the effectiveness of data visualization design. Studies commonly use completion time \cite{08rela,09rela} and error rate \cite{010rela,011rela,012rela} to measure user's efficiency and accuracy in performing tasks. Subjective evaluations are also important for assessing user experience. The NASA Task Load Index (NASA TLX) is widely used for assessing workload \cite{013rela}. Additionally, user satisfaction questionnaires offer direct feedback and emotional responses regarding the visualization tools \cite{014rela}.

\section{Study Design and Methodology}
\subsection{Sequence of Experimental Phases}
The study consisted of three distinct phases of evaluations.

The first phase focused on \textbf{node parameter optimization}, which involved a controlled experiment aimed at determining the optimal combination of node width and node quantity. Three node width conditions (narrow, medium, wide) and three node quantity conditions (low, medium, high) were evaluated. Participants performed readability and information retrieval tasks using chord diagrams with varying node width and quantity configurations. The results from this experiment informed the selection of the optimal node width and quantity settings for subsequent phases.

\begin{figure}
    \centering
    \includegraphics[width=\columnwidth]{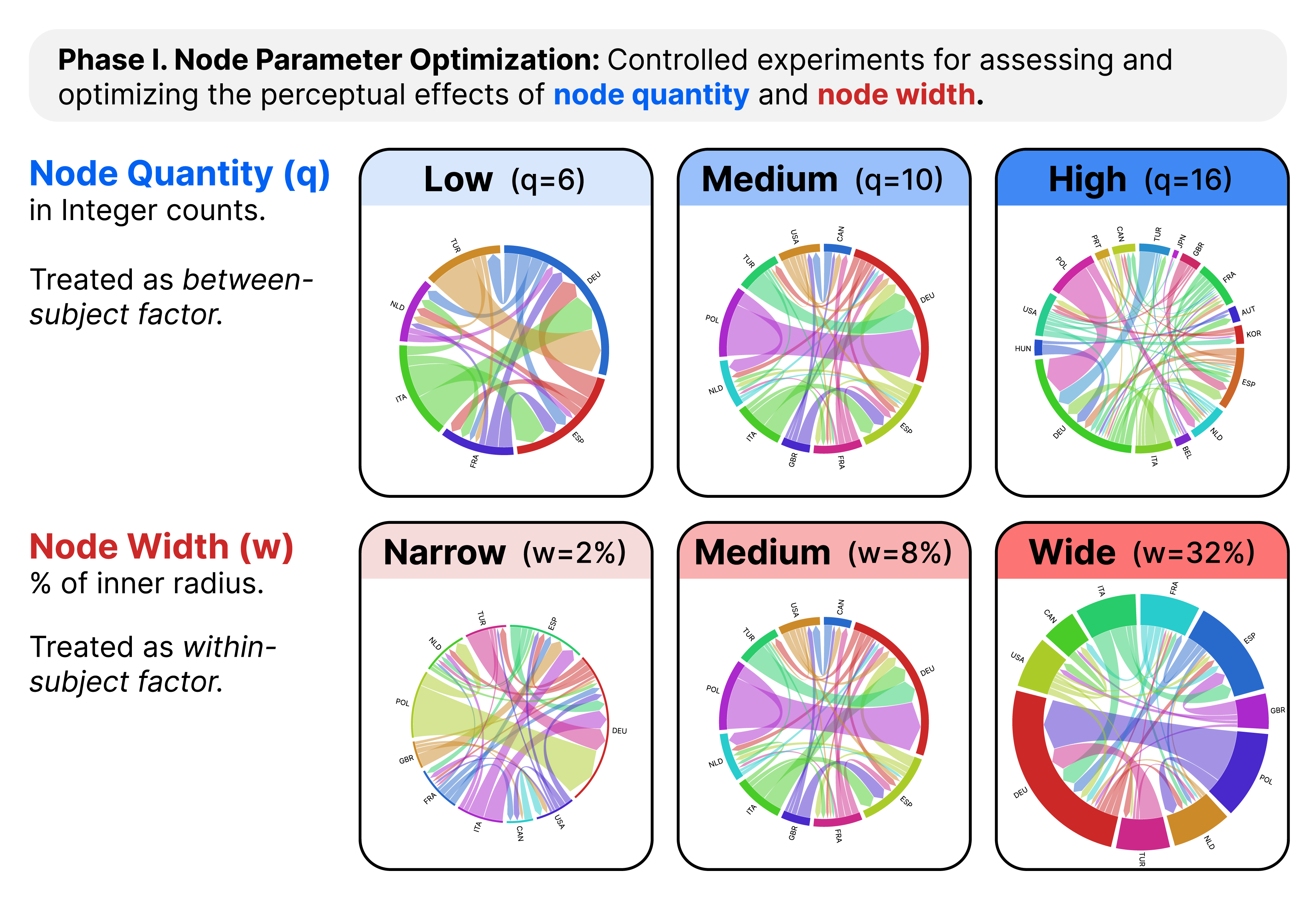}
    \caption{Experimental conditions of Phase I.}
    \Description{Figure 2 shows experimental conditions for Phase I, focusing on node parameter optimization in chord diagrams. The top row illustrates three levels of node quantity: low (q=6) with six nodes, medium (q=10) with ten nodes, and high (q=16) with sixteen nodes. The bottom row depicts three levels of node width: narrow (w=2\%) representing 2\% of the inner radius, medium (w=8\%) representing 8\% of the inner radius, and wide (w=32\%) representing 32\% of the inner radius. These conditions help in assessing and optimizing the perceptual effects of node quantity and node width. Node quantity is treated as a between-subject factor, while node width is treated as a within-subject factor.}
    \label{fig:phase1}
\end{figure}

The second phase, an \textbf{expert design review}, aimed for eliciting feedback on the design of tick marks and color gradients for chord diagrams. A diverse set of design variations were created and presented to five visualization experts. Their feedback and preference were recorded for guiding the selection of a single optimal tick mark and color gradient scheme for further evaluation in phase three.

The final phase, \textbf{design choices evaluation}, investigated the effects of the selected radial tick marks and color gradients design using a controlled user study. Participants performed tasks using chord diagrams with the optimal node width and quantity settings from the first phase, combined with the tick mark and color gradient designs chosen based on the expert review session.

\subsection{Dataset and Tasks}
For the study, we utilized a migration dataset \cite{InternationalMigrationDatabase}, a typical example of network data visualized with chord diagrams. This dataset was selected for its social relevance and practical real-world application. This dataset provides information on the number of people migrating between countries over specified periods. In the created sample chord diagrams, countries of origin and destination were represented as nodes with three-letter acronym text labels. Migration events were depicted as chord connections between nodes, and the direction of migration was shown with arrows on the arc connections. These visual elements were consistently applied throughout the study.

\begin{figure}
    \centering
    \includegraphics[width=\columnwidth]{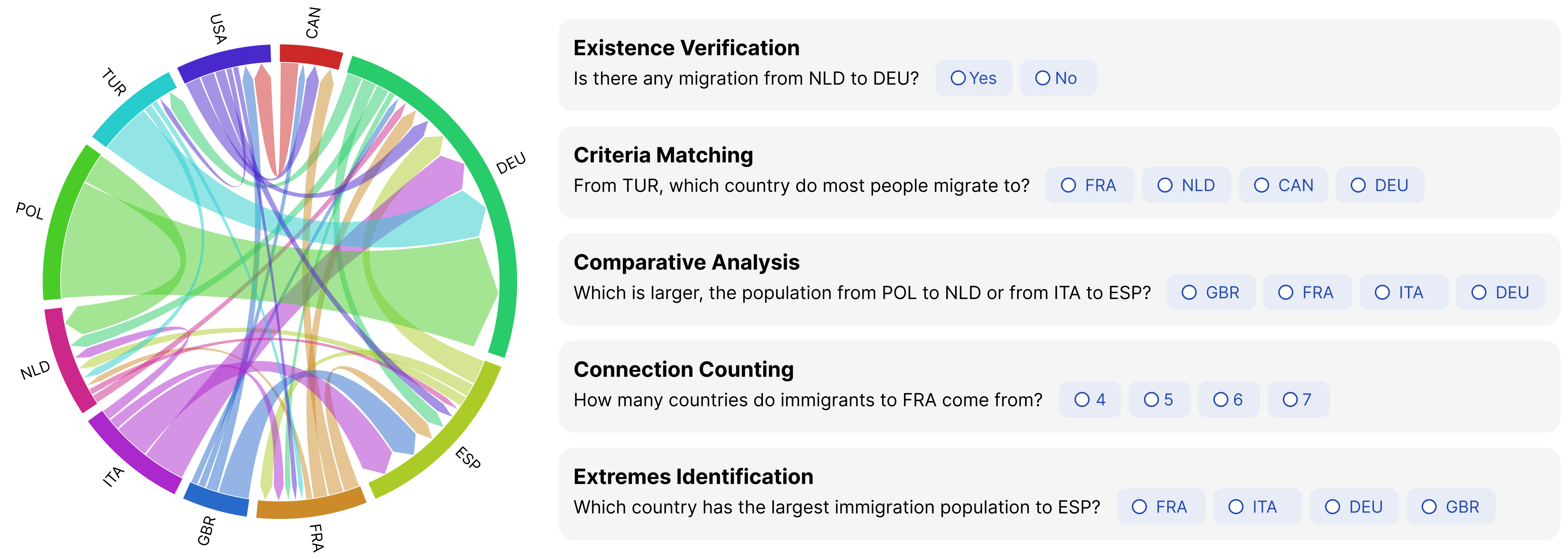}
    \caption{Summary of tasks. Left: A sample chord diagram stimuli. Right: Five types of task questions.}
    \Description{Figure 3 shows a summary of tasks using a chord diagram. The left side of the image displays a sample chord diagram with nodes representing different countries and arcs showing migration flows between them. The right side of the image lists five types of task questions: Existence Verification (e.g., Is there any migration from NLD to DEU?), Criteria Matching (e.g., From TUR, which country do most people migrate to?), Comparative Analysis (e.g., Which is larger, the population from POL to NLD or from ITA to ESP?), Connection Counting (e.g., How many countries do immigrants to FRA come from?), and Extremes Identification (e.g., Which country has the largest immigration population to ESP?).}
    \label{fig:task}
\end{figure}

We assessed the chord diagram designs using five tasks that represent typical analysis goals for network data (\autoref{fig:task}), adapted from Gutwin et al.’s study \cite{Gutwin:2023:SFC}. Tasks ranged from basic retrievals to complex comparisons, presented in a counter-balanced order to mitigate order effects.

\begin{itemize}[leftmargin=2em]
\item \textbf{Existence verification}: Identifying if a specific node or connection existed. For example, determining whether a connection existed between the United States (USA) and Canada (CAN).
\item \textbf{Criteria matching}: Identifying a node or connection that matched a specific criterion, such as finding the country with the most incoming migration.
\item \textbf{Comparative analysis}: Comparing two elements, such as determining which connection represented a larger migration flow between two countries.
\item \textbf{Connection counting}: Determining the number of incoming or outgoing connections associated with a particular node.
\item \textbf{Extremes identification}: Identifying the nodes or connections with the maximal or minimal quantities, such as the country with the highest or the lowest net immigration.
\end{itemize}

\subsection{Chord Diagram Generation and Variation}

\textbf{Baseline Chord Diagram Template.}
We used a consistent baseline template in D3.js for all diagrams, following standard conventions with color encoding for countries and directional arrows for migration flows \cite{Nguyen2012CircoSonicAS}. To mitigate learning effects, the radial arrangement of the nodes was randomized across tasks. The baseline diagram was modified to create different variants for experimental conditioning.

\vspace{0.5em}
\noindent
\textbf{Varying Node Parameters.}
% The number of nodes significantly impacts its complexity and information load contained within a chord diagram. Gutwin et al. \cite{relatedwork_6_gutwin2023showing} experimented with chord diagrams containing 8-15 nodes and defined this range as medium data size. However, there is limited classification on how the node quantity affects visualization complexity. 
The number of nodes can impact a chord diagram's complexity. To optimize the quantity of nodes, we defined three experimental conditions:  \textit{low (6 nodes)}, \textit{medium (10 nodes)}, and \textit{high (16 nodes)} (\autoref{fig:phase1} top row). The number of countries in the dataset was varied based on the baseline template while keeping other parameters identical.

Similarly, we categorized three levels of node widths: \textit{narrow}, \textit{medium} and \textit{wide} (\autoref{fig:phase1} bottom row). These widths were quantified in increments relative to the inner radius of the chord diagram's circular outline. The \textit{narrow} condition was set as 2\% of the inner radius, the \textit{medium} condition was set as 8\%, and the \textit{wide} condition was set to 32\%. These increments, set at four-fold increases, were designed to create clear visual distinctions among the levels. In our experiments, node quantity was a between-subject factor, while node width was a within-subject factor.

\begin{figure}
    \centering
    \includegraphics[width=\columnwidth]{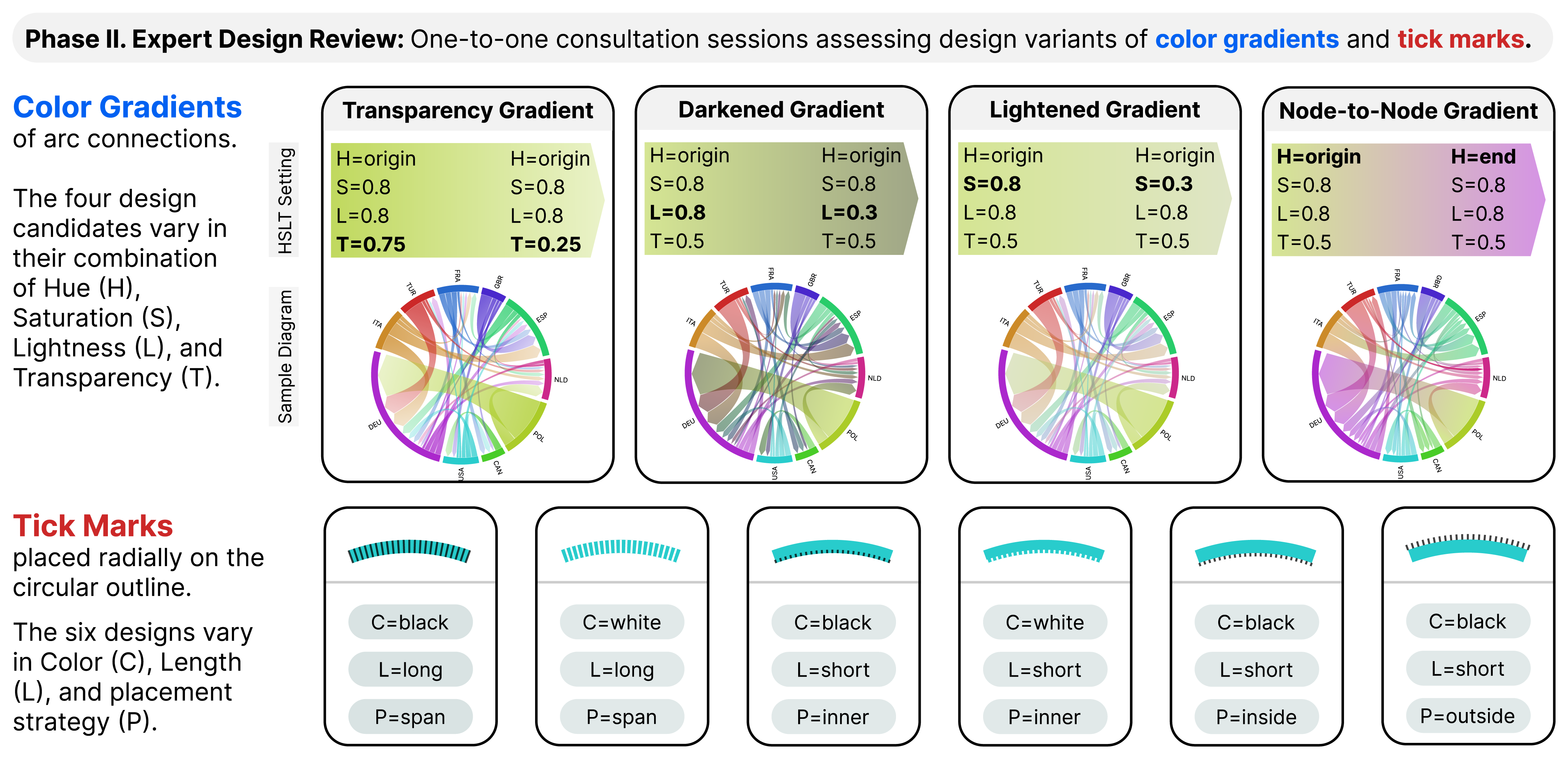}
    \caption{Experimental Condition of Phase II.}
    \Description{Figure 4 shows Phase II of the Expert Design Review, evaluating design variants of color gradients and tick marks in chord diagrams. The top section focuses on color gradients of arc connections, displaying four design candidates that vary in their combination of Hue (H), Saturation (S), Lightness (L), and Transparency (T). The variants include Transparency Gradient, Darkened Gradient, Lightened Gradient, and Node-to-Node Gradient. The bottom section addresses tick marks placed radially on the circular outline, with six design variations in Color (C), Length (L), and Placement strategy (P): C=black, L=long, P=span; C=white, L=long, P=span; C=black, L=short, P=inner; C=white, L=short, P=inner; C=black, L=short, P=inside; C=black, L=short, P=outside.}
    \label{fig:phase2}
\end{figure}

\vspace{0.5em}
\noindent
\textbf{Varying Color Gradients.}
To assess how color gradients impact the interpretability of chord diagrams, we created four gradient variations (\autoref{fig:phase2} top row) for expert review. 

\noindent
\includegraphics[height=0.8em]{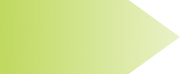} The \textit{transparency gradient} linearly adjusts the opacity of arc connections to indicate direction. As the data flows from the origin to the destination, the opacity faded from 75\% to 50\%.

\noindent
\includegraphics[height=0.8em]{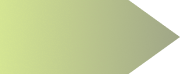} In \textit{darkened gradient}, the color of the arc connections transitions to a darker shade in lightness as the flow progresses from the origin to the destination.

\noindent
\includegraphics[height=0.8em]{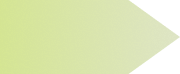} The \textit{lightened gradient} involves gradually lightening the color saturation of the arc connections as the flow moves from the origin to the destination.

\noindent
\includegraphics[height=0.8em]{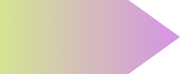} The \textit{node-to-node gradient} creates a linear color transition from the origin to the destination, with each arc starting with the origin node's color and gradually blending into the destination node's color.

\vspace{0.5em}
\noindent
\textbf{Adding Radial Tick Marks.}
To evaluate data comparison and value reading, we integrated various tick marks into the circular outline of our baseline chord diagram template. We developed six tick mark designs (\autoref{fig:phase2} bottom row), differing in color, length, and placement. The tick marks were designed in black and white, for providing contrast against the node segments. Some designs spanned the full node width, while others were shorter for visual subtlety. The tick marks were also placed in several different ways: inside the circular outline, along the inner edge of the circular outline, superimposed to span the node segments, along the outer edge of the circular outline, or at the outside of the circular outline.

\subsection{Evaluation Methods}
We adopted a hybrid evaluation methodology combining objective and subjective data. The study was approved by the university ethics committee prior to its conduct.

\vspace{0.5em}
\noindent
\textbf{Controlled Experiments in Phase I \& III.}
In Phases I (\autoref{sec:expI:node}) and III (\autoref{sec:expIII:choices}), quantitative questionnaires were designed to evaluate participants’ viewing experiences with different chord diagram designs. The questionnaires consisted of three parts: 

\begin{itemize}[leftmargin=2em]
    \item \textbf{Practice Set:} Served as a training set and introduced the format of the questionnaire with a simplified sample chord diagram (10 nodes and 11 chords). Each page displayed a sample chord diagram at the top, followed by a multiple-choice question. Participants were required to answer correctly to proceed.
    \item \textbf{Objective Performance:} The second section extended the practice set to measure objective performance with various chord diagram stimuli. Five questions per task per experimental condition were presented. Performance metrics included the time taken to complete each question (from when it appeared on the screen to when the participant confirmed their answer correctly) and the counts of error occurrences.
    \item \textbf{Subjective Experience:} 
     Participants completed the NASA Task Load Index (TLX) to and provided rankings and feedback on the diagram design based on ease of information retrieval, accuracy, and overall preference. 
     %Lastly, an optional open-ended question allowed participants to provide feedback on their experience with the chord diagrams.
\end{itemize}

\noindent
\textbf{Qualitative Design Review in Phase II.}
Phase II involved qualitative consultation sessions with visualization experts (E1 to E5). One-on-one interviews, conducted in the experts’ native languages, lasted 10-15 minutes each. The open-ended interviews began with a brief study overview, followed by a presentation of six design candidates. Subsequently, experts discussed the potential impact of these designs on the five information tasks and concluded by selecting one most-preferred color gradients and tick mark designs, respectively.

% Together, these multi-staged hybrid evaluation techniques provided a comprehensive understanding of chord diagram designs.

\section{I. Node Parameter Optimization}
\label{sec:expI:node}
In phase one, we recruited 112 participants from the university. After discarding 22 responses due to outlying completion times (over 15 minutes or less than one second per question), we had 90 valid responses (age: 25 $\pm$ 6; gender distribution: 52 females, 35 males, 2 others, 1 undisclosed). Among these participants, 6 were very familiar with data visualizations, 29 were familiar, 39 had some knowledge, and 16 were completely unfamiliar. 

Completion time and error occurrences for each question were collected, aggregated and averaged by tasks, resulting in 1,350 objective measures (30 participants $\times$ 3 node quantities $\times$ 3 node widths $\times$ 5 tasks). Additionally, we gathered 270 sets of subjective measures from the NASA TLX and preference rankings (30 participants $\times$ 3 node quantities $\times$ 3 node widths).  These data were analyzed with mixed ANOVA and Pearson Chi-Squared tests, respectively, with detailed results available in Appendix \ref{sec:app:phaseI}.

\begin{figure*}
    \centering
    \includegraphics[width=\linewidth]{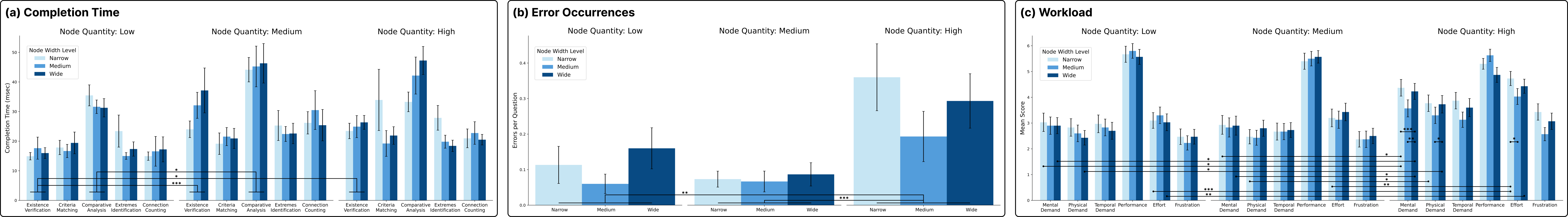}
    \caption{Phase I Results: Average completion time, error occurrences and subjective ratings of workload of different node quantities and widths across tasks. Significance levels: $p < .05$(*), $p < .01$(**), and $p < .001$(***).}
    \Description{Figure 5 shows Phase I results with three panels. Panel (a) presents average completion time for tasks with different node quantities (Low, Medium, High) and widths (Narrow, Medium, Wide). Lower node quantities resulted in shorter completion times for existence verification and comparative analysis tasks. Panel (b) displays error occurrences, with medium node quantity resulting in the fewest errors. Both low and medium node quantities had fewer errors than high node quantity. Panel (c) shows subjective workload ratings, indicating lower mental and physical demands with fewer nodes and medium node widths, and lower perceived effort with low node quantity and narrow width. Medium node width reduced frustration and overall workload with high node quantity.}
    \label{fig:phaseIResult}
\end{figure*}

\subsection{Node Quantity Impacted Performance Metrics}

\vspace{0.5em}
\noindent
\textbf{Fewer nodes led to shorter completion time in existence verification and comparative analysis.}
As shown in \autoref{fig:phaseIResult} (left), for existence verification tasks, low node quantity resulted in the shortest completion times compared to medium ($p <$ 0.001) and high node quantities ($p =$ 0.024). Similarly, for comparative analysis, low node quantity resulted in significantly shorter completion times than medium node quantity ($p = 0.042$). No significant effects in completion time were found for criteria matching, extremes identification, or connection counting across different node quantities. 

\vspace{0.5em}
\noindent
\textbf{Medium node quantity resulted in fewer errors.}
As shown in \autoref{fig:phaseIResult} (middle), node quantity significantly affected error occurrences ($F_{(2, 87)}$ = 8.246, $p <$ 0.001). Both low and medium node quantities had significantly lower error occurrences compared to high node quantity ($p = 0.007$ and $p <$ 0.001, respectively). Medium node quantity had the lowest mean error occurrences, though the difference between low and medium was not statistically significant.

\subsection{Interaction Effects Between Node Width and Quantity in Subjective Experience}
Statistically significant interaction effects between node quantities and node widths were identified for mental demand ($F_{(4, 174)} = 2.564, p = 0.040$), physical demand ($F_{(4, 174)} = 2.670, p = 0.034$), and perceived effort ($F_{(4, 174)} = 2.540, p = 0.042$) (\autoref{fig:phaseIResult} (right)).

\vspace{0.5em}
\noindent
\textbf{Fewer Nodes Led to Lower Mental and Physical Demands.}
For narrow nodes, significantly lower mental demands were observed between low and high quantities ($p = 0.025$), as well as between medium and high quantities ($p = 0.014$). Besides, medium quantity showed a significantly lower physical demand to high quantity ($p = 0.013$). For wide nodes, low quantity also led to significantly lower mental demand ($p = 0.017$) and physical demand ($p = 0.012$) than high quantity. Similarly, medium quantity resulted in lower mental demand compared with high quantity ($p = 0.017$).

\vspace{0.5em}
\noindent
\textbf{Lower Node Quantity Reduced Perceived Effort.}
Low node quantity with narrow node width led to significantly lower perceived effort compared to high node quantity ($p < 0.001$) and medium node quantity ($p = 0.002$). For wide node width, low node quantity resulted in significantly lower perceived effort compared to high node quantity ($p = 0.007$).

\vspace{0.5em}
\noindent
\textbf{Medium Node Widths Resulted in Less Frustration.}
Node width made a significant difference in frustration ($p = 0.019$). Post hoc analysis revealed medium width led to significantly lower frustration compared to narrow nodes ($p = 0.016$).

\vspace{0.5em}
\noindent
\textbf{Medium Node Widths Reduced Mental, Physical Demands and Effort with High Node Quantity.}
For high node quantity, medium node width consistently resulted in lower workload. Medium width led to significantly lower mental demand compared to narrow ($p < 0.001$) and wide ($p = 0.005$). Similarly, medium width reduced physical demand for high node quantity compared to wide nodes ($p = 0.043$). Additionally, medium width resulted in significantly lower perceived effort when compared to narrow nodes ($p = 0.014$).

\subsection{Preferences and Qualitative Feedback}
Over 56\% of participants preferred the medium width for better speed (56.7\%, N = 51), better accuracy (60\%, N = 54), and overall perception (62.2\%, N = 56). No significant association was found between the preference rankings of node quantities and widths. The participants mentioned that the node width was particularly important in comparative tasks, especially when the flows were of similar size. One participant commented, ``It is tough to compare the smaller ones (flows); mostly it’s a guess which one is larger''. Participants with more nodes felt that the many links in the chord diagrams increased their visual burden. One said, ``If a country has too many migration lines, it visually becomes a bit chaotic.''

\revised{In summary, fewer nodes led to faster completion times for tasks like existence verification and comparative analysis, showing that simplifying visual complexity improves information retrieval speeds without losing accuracy. Medium node quantities reduced errors, suggesting a good balance between detail for accurate analysis and reduced cognitive load.
Interactions between node quantities and widths also affected perceived workload, highlighting how these elements together influence user experience. Overall, medium node width was favored for its speed, accuracy, and overall perception.} Therefore, we proceeded with medium node width and medium node quantity in the subsequent phases.

\section{II. Expert Design Review}
\label{sec:expII:review}
We invited five visualization experts to provide design insights on color gradients and tick marks (Appendix \autoref{fig:A2}). 

\vspace{0.5em}
\noindent
\textbf{Evaluation of Color Gradients.}
Three experts favored transparency gradient \includegraphics[height=0.6em]{Icon/TransparencyGradient.png}  for its effectiveness across various tasks. They highlighted that changes in transparency maintained a high level of stylistic consistency and visual effects. E1 noted, ``The transparency gradient just looks overall brighter''.
%They found this gradient easy to read and visually appealing.

The darkened gradient \includegraphics[height=0.6em]{Icon/DarkenedGradient.png} was less favored due to its dimmer appearance when blending multiple colors. The lightened hue gradient \includegraphics[height=0.6em]{Icon/LightenedGradient.png} received mixed reviews. While some experts appreciated its visual appeal, others were concerned about its reduced saturation in dense diagrams. The node-to-node gradient \includegraphics[height=0.6em]{Icon/Node2NodeGradient.png} was less favored because of the potential visual complexity when multiple colors were involved. Overall, the consensus leaned towards transparency gradients for their clarity and readability.

\vspace{0.5em}
\noindent
\textbf{Evaluation of Tick Mark Designs.}
All experts agreed that the tick mark colors should have high contrast against the background to ensure readability. White tick marks \includegraphics[height=1em]{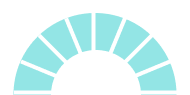} \includegraphics[height=1em]{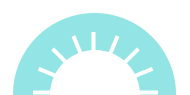} were thought to provide better legibility against blue and purple backgrounds.

On the other hand, opinions diverged on the placement strategies of the tick marks. Two experts preferred tick marks spanning the entire node width for consistent visual cues \includegraphics[height=1em]{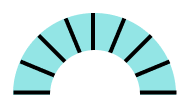} \includegraphics[height=1em]{Icon/slice_b.png}. However, one expert criticized this approach because this placement ``made the nodes appear disjointed'' (E3). One expert applauded the placement approach inside the circular outline \includegraphics[height=1em]{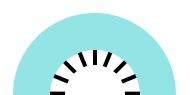}, arguing that placing the tick marks in proximity to the arc connections ``reduced the effort required during comparison,'' (E2), yet two experts criticized it for ``occupying additional space and overlapping with some arc connections.'' Two experts preferred white tick marks along the inner edge \includegraphics[height=1em]{Icon/slice_e.png}, which combines the advantages of proximity to the chords while not invading the space of the chords themselves. None of the experts favored placing the tick marks outside the circular outline \includegraphics[height=1em]{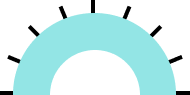} due to concerns that it ``occupies extra space and is too distant from the chords upon comparison'' (E1, E2, E4). After synthesizing their opinions, we finalized the design choices as white tick marks along the inner edge of the circular outline. This combination was selected for clarity, contrast and readability. 

\section{III. Design Choices Evaluation}
\label{sec:expIII:choices}
\begin{figure}
    \centering
    \includegraphics[width=\columnwidth]{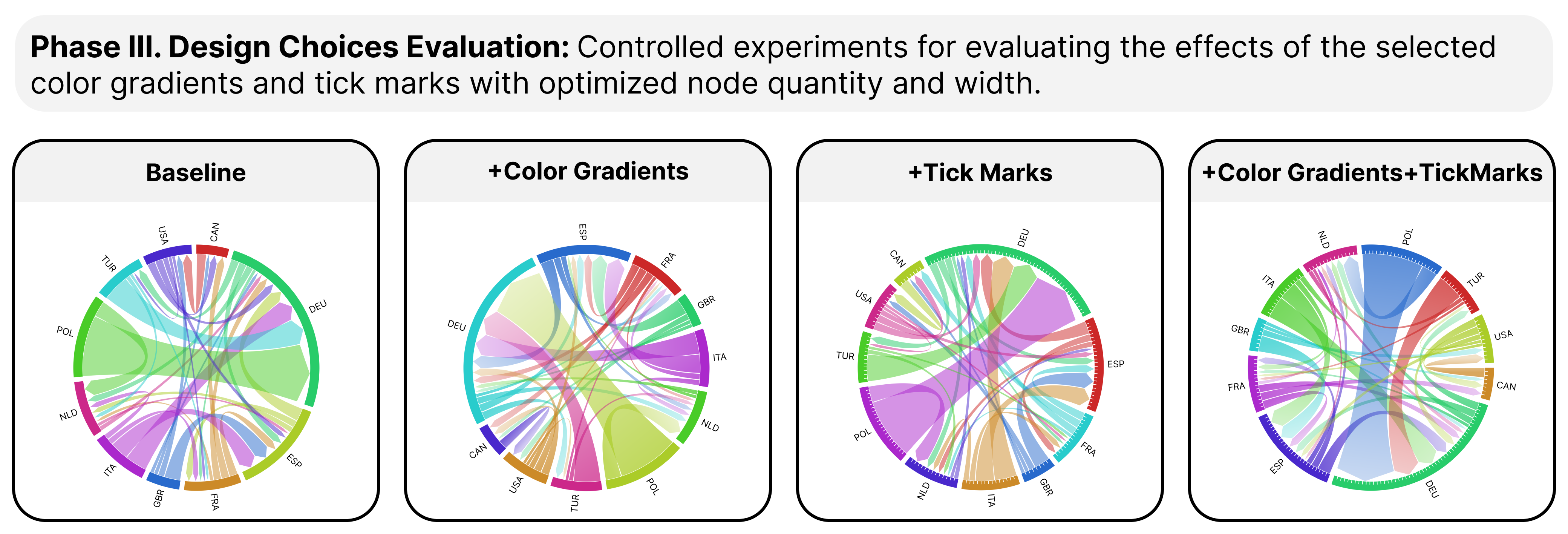}
    \caption{Experimental conditions of phase III.}
    \Description{Figure 6 shows the experimental conditions of Phase III. There are four chord diagrams displayed: Baseline, Baseline with Color Gradients, Baseline with Tick Marks, and Baseline with both Color Gradients and Tick Marks. The Baseline diagram is a chord diagram template with 10 nodes and medium node width, serving as a reference before any design alterations. The Baseline + Color Gradient diagram has arc connections altered with a transparency gradient. The Baseline + Tick Marks diagram includes white tick marks added to the nodes, extending outward from the inner edge to one-third of the node length. The Baseline + Color Gradients + Tick Marks diagram combines the color gradients and tick mark designs with the baseline.}
    \label{fig:phase3}
\end{figure}

Building on findings from Phase I and II, we assessed chord diagram perception under four design configurations (\autoref{fig:phase3}):

\begin{itemize}[leftmargin=2em]
    \item \textbf{Baseline:} Chord diagram template with 10 nodes and medium node width, serving as reference before any design alterations.
    \item \textbf{Baseline + Color Gradient:} Arc connections were altered with a transparency gradient.
    \item \textbf{Baseline + Tick Marks:} White tick marks were added onto the nodes, extending outward from the inner edge of the nodes to $\frac{1}{3}$ of node length.
    \item \textbf{Baseline + Color Gradients + Tick Marks:} Combined color gradients and tick mark designs with the baseline.
\end{itemize}

We recruited 24 participants from the university (age: 24 $\pm$ 2; gender distribution: 13 females, 11 males), all of whom had normal visual acuity and varying experience with visualization (1 very familiar, 10 familiar, 13 somewhat knowledgeable). %Regarding their knowledge of chord diagrams, 4 were familiar, 12 had some knowledge, and 8 were completely unfamiliar. 
In total, we collected 480 sets of task completion time and error occurrences data (24 participants $\times$ 4 conditions $\times$ 5 tasks), and 96 sets of subjective measures on workload and preference (24 participants $\times$ 4 conditions). Additionally, we compiled interview records totaling 63 minutes. \revised{While we acknowledge the relatively small sample size, this number was chosen to balance statistical power with the ability to conduct in-depth interviews for rich qualitative insights into user experience.} For data analysis, we conducted the Friedman test due to non-normality of the dependent variables across the four conditions.

\begin{figure*}
    \centering
    \includegraphics[width=\linewidth]{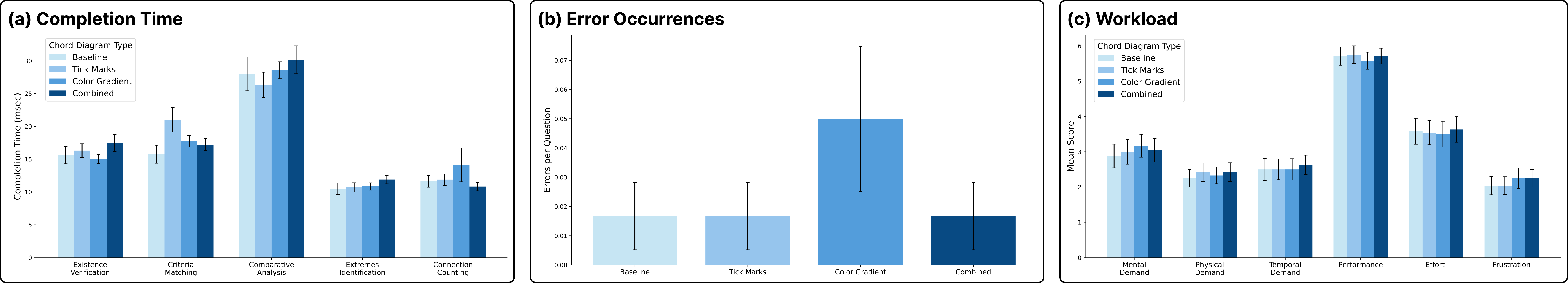}
    \caption{Phase III Results: Average completion time, error occurrences and subjective ratings of workload of different design choices across tasks. Significance levels: $p < .05$(*), $p < .01$(**), and $p < .001$(***).}
    \Description{Figure 7 shows Phase III results with three panels. Panel (a) presents average completion time for tasks under different chord diagram types: Baseline, Tick Marks, Color Gradient, and Combined. No statistically significant differences were found in completion time among the four conditions. Panel (b) shows error occurrences, with the highest error rate in the Color Gradient condition, which decreased in the Combined condition, suggesting tick marks helped reduce errors. Panel (c) displays subjective workload ratings, including mental demand, physical demand, temporal demand, performance, effort, and frustration, with no significant differences among the conditions.}
    \label{fig:phaseIIIResult}
\end{figure*}

\revised{Despite rigorous experimental design, the Phase III results (\autoref{fig:phaseIIIResult}) did not reveal statistically significant differences in completion time, error occurrences, or workload among the four conditions ($p > 0.1$) (Appendix \ref{sec:app:phaseIII}). This outcome may suggest several key points about the experimental design and metric selection.}

\vspace{0.5em}
\noindent
\revised{
\textbf{Factors Influencing Experimental Sensitivity.} The lack of statistically significant results could be due to several factors. Firstly, the sensitivity and specificity of the chosen metrics, particularly completion time and error rates, may not have effectively captured the subtle effects of design changes on user performance. This metric granularity may have limited the chance to accurately reflect the cognitive processes involved in interpreting chord diagrams. Additionally, the visual tasks used may not have been distinct enough to show noticeable performance differences among the design conditions. Future studies might benefit from either more pronounced design changes or a variety of tasks designed to highlight specific design features. Lastly, the homogeneity of the participant pool, marked by similar backgrounds as students and age group, potentially diluted the observable impact of the design modifications. A more diverse group of participants could likely yield more significant differences to enhance the generalizability and sensitivity of the results.}

\vspace{0.2em}
Apart from the broader challenges in experimental sensitivity, several interesting trends emerged in the observation of key metrics.

\vspace{0.5em}
\noindent
\textbf{Task-Specific Completion Times.}
The completion times varied across different tasks in a consistent manner. Notably, the \textit{comparative analysis} tasks consistently required the longest completion times, potentially indicating that the inherent complexities of specific tasks dictate the time required for completion. This observation is consistent with the time recorded in Phase I, suggesting that complex tasks demand more time irrespective of design changes.

\vspace{0.5em}
\noindent
\textbf{Error Occurrences Across Designs.}
The error occurrences peaked in the \textit{color gradient} condition. However, after the addition of tick marks in the \textit{combined} condition, the error occurrences returned to a lower level (\autoref{fig:phaseIIIResult}). This reduction may suggest that tick marks potentially provided visual cues that aided in the interpretation of the gradients, which almost rescued the initial increase in errors.

\vspace{0.5em}
\noindent
\textbf{Preferences and Qualitative Feedback.} 
Feedback on adding tick marks was generally positive. About 37.5\% (N=9) of participants believed that chord diagrams with tick marks helped them find answers more quickly and accurately. Many participants (N=14) noted that tick marks allowed for more accurate comparisons of chord widths. 

Some participants (N=6) noted areas for improvement. One mentioned that reading values from tick marks was not straightforward, suggesting, ``It would be better to directly label the numeric values instead of making me count them.'' (P14). There were also issues with precision: ``Some chords are narrower than the smallest tick mark, making it hard to use the marks to speed up comparisons'' (P20). Additionally, tick marks did not always align perfectly with the chords, ``Not all tick marks start at one end of the chord, which might introduce errors in comparisons'' (P4).

A few participants (N=4) criticized tick marks for increasing cognitive load and distraction. One mentioned that ``When not comparing tasks, tick marks distract my attention; having the option to display tick marks would be better''(P10).

Regarding the transparency gradients, some participants (N=4) did not notice their presence and questioned the difference between graphs with and without these gradients. Only four users explicitly expressed the benefits of transparency gradients for distinguishing flow direction, describing it as ``a psychological hint'' of directionality (P1). However, the majority of users (N=13) preferred a simpler design. As one participant (P11) pointed out, ``Similar colors become indistinguishable after transparency gradients.''

\section{Limitations of The Study}

\textbf{Challenges in Color Gradients and Tick Marks Design.}
In the experiment, many users either did not notice the gradient colors or found them unhelpful. This may be due to the gradient scheme not being visually prominent enough. In future design iterations, more diverging and easily distinguishable gradient schemes should be considered. As for the tick marks, the added detail from tick marks might lead to unnecessary cognitive burdens in simpler tasks. Tick marks should be used selectively according to task complexity.

\vspace{0.5em}
\noindent
\textbf{Difficulty in Tasks and Questions.}
This study utilized five tasks to evaluate chord diagram designs. However, one particular task, comparative analysis, consistently took longer across different conditions. This indicates that our current tasks might not fully encompass the spectrum of difficulty or adequately represent all potential user interactions. To address this, future research should expand the range of tasks to more effectively assess design variants across varied difficulty levels, particularly those involving comparative numerical analysis. These questions are likely to place greater demands on the precision of tick marks and the visual clarity of the chord diagrams.

\vspace{0.5em}
\noindent
\revised{
\textbf{Dataset and Generalizability.} 
In this study, we utilized a specific migration dataset for its real-life application as a type of network data. It is important to note that the findings may have limited generalizability due to the dataset’s specific characteristics. Future studies could enhance the applicability of these results by incorporating a diverse range of datasets, including those with varying sizes, and complexities, thereby strengthening the validity of the recommendations for broader use in network data visualization.
}

\section{Practical Guidelines}
\revised{
\textbf{Finding the Sweet Spot: Optimizing Node Width and Quantity.} Our findings suggest that using a medium number of nodes ($n=10$) and medium widths generally offers the best readability and user satisfaction under the current experimental conditions. While the exact number may vary, the rule of thumb is to strike a balance between providing sufficient detail and avoiding visual clutter.
When working with larger datasets, designers can consider implementing selective filtering to maintain this optimal balance. For instance, filtering functionalities enable users to focus on a selected subset of nodes at a time. An overview-detail approach can also be effective, presenting a full chord diagram for context alongside a more detailed view of selected nodes, where users can interpret data up close.
% Additionally, dynamically adjusting node widths can improve focus and clarity. For example, increasing a node’s width when hovered over can help highlight important information, making the diagram easier to navigate and understand.
}

\vspace{0.5em}
\noindent
\revised{
\textbf{Navigating Directionality: Customizing Color Gradients and Beyond.} 
When approaching the directional representation of flows, designers can consider using color gradients or alternative methods with flexibility and customization. For example, the study tested different styles of color gradients but not their mapping polarity. Experimenting with whether more vibrant colors are placed at the source or destination node could enhance intuitive understanding of data flow direction. Offering users the ability to customize the direction-color gradient mapping can also be beneficial. This allows users to adjust the visualization to fit their individual preferences.
}

\vspace{0.5em}
\noindent
\revised{
\textbf{Making Comparison Easier: Tick Marks Where It Counts.}  
Tick marks enrich the user’s subjective experience by offering clear visual references. These markers are especially useful in scenarios that demand data comparison. For tasks that require a broad overview or coarse comparisons, incorporating tick marks can help viewers quickly gauge differences between values. However, when precise data values are crucial and exact quantities are statistically relevant, designers can display numerical values directly on the diagram. Designers can explore different techniques for placing the tick marks or numerical values within a chord diagram: interactive elements such as hover tooltips, toggle switches or zooming features can be implemented to selectively reveal specific data when a user focuses on a segment or connection.
}

\vspace{0.5em}
\noindent
\revised{
\textbf{One Size Does Not Fit All: CHORDinating Design Elements.}  
The interaction effects observed between node quantity and width, as well as the combined influence of tick marks and color gradients, demonstrate that different design elements can interact in complex ways that influence overall effectiveness of visualizations. Instead of simply layering design strategies, testing how different design elements work together holistically can reveal insights into how they influence user perception and task performance. For instance, while medium node widths generally improve readability, their effectiveness can be contingent on the node quantity present. Similarly, the benefits of color gradients in indicating direction can be amplified or obscured by how tick marks are implemented. Therefore, designers should conduct regular user testings and use iterative design processes to fine-tune how elements like node size, tick marks, and color gradients combine.
}

\section{Conclusion}

This study explored the key elements of chord diagram design and their impact on user perception and information acquisition. Through three experimental phases, we assessed the effects of node width and quantity, as well as radial tick marks and color gradients. Node width did not significantly affect performance metrics, but impacted subjective experiences. Medium node width was preferred by the majority of users. Increasing the number of nodes extended task completion time and increased error occurrences, especially in tasks involving comparison and existence verification. Tick marks improved the perceived accuracy of data interpretation, while color gradients, despite aiming to enhance understanding of data flow, had limited practical effects. Future research should optimize these to improve the usability and effectiveness of chord diagrams.

%---
\begin{acks}
This work was supported by the National Natural Science Foundation of China (Grant No. 62272396).
\end{acks}

\bibliographystyle{ACM-Reference-Format}
\bibliography{references}

\appendix
\setcounter{figure}{0}
\onecolumn

\section{Tick Marks Design Candidates in Phase II}
\begin{figure}[h]
    \includegraphics[width=0.9\linewidth]{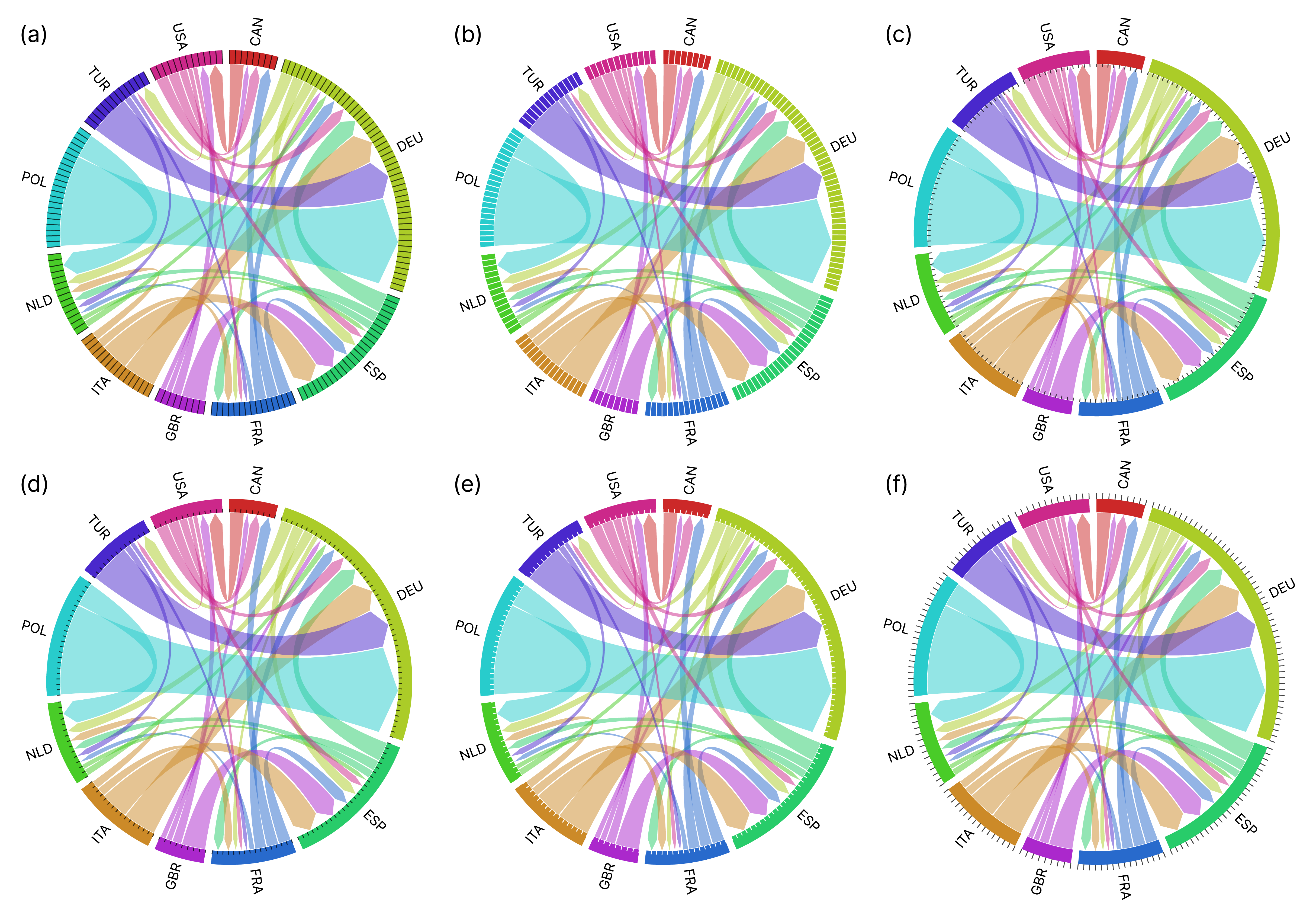}
    \caption{Six design candidates of tick marks that were presented to the experts for review.}
    \Description{Figure 1 shows six design candidates of tick marks presented to the experts for review. Each subfigure (a) through (f) displays a chord diagram with different configurations of tick marks.
(a) Tick marks are black, long, and span the entire outer edge of the nodes.
(b) Tick marks are white, long, and span the entire outer edge of the nodes.
(c) Tick marks are black, short, and positioned at the inner edge of the nodes.
(d) Tick marks are white, short, and positioned at the inner edge of the nodes.
(e) Tick marks are black, short, and placed inside the nodes.
(f) Tick marks are black, short, and placed outside the nodes.
}
    \label{fig:A2}
\end{figure}

\newpage
\section{Summary of Statistical Results from Phase I}
\label{sec:app:phaseI}

\subsection{Completion Time For Different Node Quantities and Node Widths}
\label{sec:Result_TIME_1}

\begin{table}[ht]
    \centering
    % \caption{Experiment I Completion Time (sec)}
    \resizebox{\textwidth}{!}{
    \begin{tabular}{clrrrrrrrrrrrr}
        \toprule
        & & \multicolumn{3}{c}{Low Node Quantity} & \multicolumn{3}{c}{Medium Node Quantity} & \multicolumn{3}{c}{High Node Quantity} & \multicolumn{3}{l}{ANOVA} \\
        \cmidrule(lr){3-5} \cmidrule(lr){6-8} \cmidrule(lr){9-11} \cmidrule(lr){12-14}
        & & \multicolumn{1}{c}{Narrow} & \multicolumn{1}{c}{Medium} & \multicolumn{1}{c}{Wide} & \multicolumn{1}{c}{Narrow} & \multicolumn{1}{c}{Medium} & \multicolumn{1}{c}{Wide} & \multicolumn{1}{c}{Narrow} & \multicolumn{1}{c}{Medium} & \multicolumn{1}{c}{Wide} & \multicolumn{1}{l}{Interaction} & \multicolumn{1}{l}{Node Quantity} & \multicolumn{1}{l}{Node Width}  \\
        \cmidrule(lr){1-11} \cmidrule(lr){12-14} 
        Existence & Mean & 14.915 & 17.638 & 15.981 & 24.011 & 32.109 & 37.141 & 23.444 & 24.912 & 26.353 & $F_{\left(3.55, 154.439\right)}$ = 0.792 & $F_{\left(2, 87\right)}$ = 10.908 & $F_{\left(1.775, 154.439\right)} = 1.871$   \\
        Verification & SE & 1.244 & 3.727 & 1.819 & 2.805 & 4.391 & 7.578 & 2.609 & 3.722 & 2.340 & $p$ = 0.519 & \textbf{\bm{$p<$} 0.001} & $p$ = 0.162 \\
        \addlinespace
        Criteria& Mean & 17.865 & 16.685 & 19.417 & 19.147 & 21.543 & 20.924 & 33.930 & 19.219 & 21.905 & $F_{\left(3.092, 134.489\right)}$ = 1.083 & $F_{\left(2, 87\right)}$ = 2.010 & $F_{\left(1.546, 134.489\right)}$ = 0.687  \\
        Matching& SE & 2.414 & 2.232& 3.665 & 3.628 & 3.052 & 3.394 & 10.342 & 4.350 & 2.967 & $p$ = 0.359 & $p$ = 0.140 & $p$ = 0.469 \\
        \addlinespace
        Comparative& Mean & 35.466 & 31.621 & 31.279 & 44.147 & 45.239 & 46.301 & 33.273& 42.165 & 47.250 & $F_{\left(4, 174\right)}$ = 1.377 & $F_{\left(2, 87\right)}$ = 3.232 & $F_{\left(2, 174\right)}$ = 0.732  \\
        Analysis& SE & 3.542 & 2.281 & 3.091 & 4.128& 6.871 & 6.666 & 3.318 & 6.248& 4.735 & $p$ = 0.244 & \textbf{\bm{$p$} = 0.044}& $p$ = 0.482   \\
        \addlinespace
        Extremes  & Mean & 23.404 & 14.999 & 17.351 & 25.285 & 22.422 & 22.592& 27.883 & 19.822& 18.415 & $F_{\left(3.234, 140.673\right)}$ = 0.371 & $F_{\left(2, 87\right)}$ = 1.411 & $F_{\left(1.617, 140.673\right)}$ = 3.541  \\
        Identification& SE & 5.413 & 1.131 & 2.413 & 5.066 & 2.526 & 3.451 & 4.174 & 2.190 & 1.924  & $p$ = 0.789 & $p$ = 0.249 & \textbf{\bm{$p$} = 0.041} \\
        \addlinespace
        Connection & Mean & 14.941 & 16.567 & 17.229 & 26.193 & 30.498 & 25.398 & 20.951 & 22.751& 20.507 & $F_{\left(2.568, 111.728\right)}$ = 0.165 & $F_{\left(2, 87\right)}$ = 2.685 & $F_{\left(1.284, 111.728\right)}$ = 0.327  \\
        Counting& SE & 1.416 & 5015 & 4.224 & 3.767& 6.545 & 5.298 & 3.170 & 3.810 & 1.779 & $p$ = 0.895 & $p$ = 0.074 & $p$ = 0.624  \\
        \bottomrule
    \end{tabular}    
    }
\end{table}

\begin{table}[ht]
    \centering
    \resizebox{\textwidth}{!}{
    \begin{tabular}{clrrrrrr}
        \multicolumn{8}{c}{Post hoc Analysis} \\ 
        \toprule
            ~ & ~ & Low-Medium & Low-High & Medium-High & Narrow-Medium & Narrow-Wide & Medium-Wide \\
            \midrule
            Existence & Mean Difference & -14.909 & -8.725 & 6.184 & ~ & ~ & ~ \\
            Verification & SE & 3.207 & 3.207 & 3.207 & ~ & ~ & ~ \\
            & $p$ & \textbf{\bm{$<$} 0.001} & \textbf{0.024} &0.171 & ~ & ~ & ~ \\
            \addlinespace
            Comparative  & Mean Difference & -12.440 & -8.107 & 4.333 & ~ & ~ &   \\
            Analysis & SE & 4.968 & 4.968 & 4.968 & ~ & ~ &   \\
            & $p$ & \textbf{0.042} & 0.319 & 1.000 & ~ & ~ &   \\
            \addlinespace
            Extremes & Mean Difference & ~ & ~ & ~ & 6.443 & 4.072 & -0.372  \\
            Identification & SE & ~ & ~ & ~ & 3.008 & 3.054 & 1.949  \\
            & $p$ & ~ & ~ & ~ & 0.105 & 0.150 & 1.000\\
            \bottomrule
    \end{tabular}
    }
\end{table}

\subsection{Error Occurrences For Different Node Quantities and Node Widths}
\label{sec:Result_ERROR_1}
\begin{table}[ht]
    \centering
    % \caption{Experiment I Error Occurrences}
    \resizebox{\textwidth}{!}{
    \begin{tabular}{cc}
        \begin{tabular}{clrrrrrrrrrrrr}
            \toprule
            & & \multicolumn{3}{c}{Low Node Quantity} & \multicolumn{3}{c}{Medium Node Quantity} & \multicolumn{3}{c}{High Node Quantity} & \multicolumn{3}{l}{ANOVA} \\
            \cmidrule(lr){3-5} \cmidrule(lr){6-8} \cmidrule(lr){9-11} \cmidrule(lr){12-14}
            & & \multicolumn{1}{c}{Narrow} & \multicolumn{1}{c}{Medium} & \multicolumn{1}{c}{Wide} & \multicolumn{1}{c}{Narrow} & \multicolumn{1}{c}{Medium} & \multicolumn{1}{c}{Wide} & \multicolumn{1}{c}{Narrow} & \multicolumn{1}{c}{Medium} & \multicolumn{1}{c}{Wide} & \multicolumn{1}{l}{Interaction} & \multicolumn{1}{l}{Node Quantity} & \multicolumn{1}{l}{Node Width}  \\
            \cmidrule(lr){1-11} \cmidrule(lr){12-14} 
            Average & Mean & 0.113 & 0.060 & 0.160 & 0.073 & 0.067 & 0.087 & 0.360 & 0.193 & 0.293 & $F_{\left(4, 174\right)}$ = 0.785 & $F_{\left(2, 87\right)}$ = 8.246 & $F_{\left(2, 174\right)}$ = 2.123   \\
            Error & SE & 0.052 & 0.027 & 0.057 & 0.022 & 0.029 & 0.033 & 0.094 & 0.071 & 0.077 & $p$ = 0.537 & \textbf{\bm{$p<$} 0.001} & $p$ = 0.123 \\
    
            \bottomrule
        \end{tabular}
    \\
    \\
        \begin{tabular}{clrrr}
        \multicolumn{5}{c}{Post hoc Analysis Results} \\ 
        \toprule
            ~ & ~ & Low-Medium & Low-High & Medium-High \\
            \midrule
            Average & Mean Difference &0.036 & -0.171 & -0.207\\
            Error & SE & 0.054 & 0.054 & 0.054\\
            & $p$ & 1 & \textbf{0.007}  & \textbf{\bm{$<$} 0.001}\\
            \bottomrule
        \end{tabular}
    \\        
    \end{tabular}
    }
\end{table}

\newpage
\subsection{Subjective Ratings on Workload}
\begin{table}[ht]
    \centering
    % \caption{Experiment I Workload}
    \resizebox{\textwidth}{!}{
    \begin{tabular}{cccc}
        \begin{tabular}{clrrrrrrrrrrrr}
            \toprule
            & & \multicolumn{3}{c}{Low Node Quantity} & \multicolumn{3}{c}{Medium Node Quantity} & \multicolumn{3}{c}{High Node Quantity} & \multicolumn{3}{l}{ANOVA} \\
            \cmidrule(lr){3-5} \cmidrule(lr){6-8} \cmidrule(lr){9-11} \cmidrule(lr){12-14}
            & & \multicolumn{1}{c}{Narrow} & \multicolumn{1}{c}{Medium} & \multicolumn{1}{c}{Wide} & \multicolumn{1}{c}{Narrow} & \multicolumn{1}{c}{Medium} & \multicolumn{1}{c}{Wide} & \multicolumn{1}{c}{Narrow} & \multicolumn{1}{c}{Medium} & \multicolumn{1}{c}{Wide} & \multicolumn{1}{l}{Interaction} & \multicolumn{1}{l}{Node Quantity} & \multicolumn{1}{l}{Node Width}  \\
            \cmidrule(lr){1-11} \cmidrule(lr){12-14} 
            Mental & Mean & 3.033 & 2.900 & 2.900 & 2.933 & 2.833 & 2.900 & 4.367 & 3.567 & 4.233 & $F_{\left(4, 174\right)}$ = 2.564 & %$F_{\left(2, 87\right)}$ = 4.141
            & %$F_{\left(2, 174\right)}$ = 0.357   
            \\
            Demand & SE & 0.357 & 0.337 & 0.312 & 0.365 & 0.372  & 0.369  & 0.323  & 0.331  & 0.309  & \textbf{\bm{$p =$} 0.040} & %\textbf{\bm{$p=$ 0.019}} 
            & %\textbf{\bm{$p=$ 0.008}} 
            \\
            %\midrule
            \addlinespace
            Physical& Mean & 2.833 & 2.600 & 2.433 & 2.467 & 2.433 & 2.800 & 3.767& 3.300 & 3.733 & $F_{\left(4, 174\right)}$ = 2.670 & %$F_{\left(2, 87\right)}$ = 3.951 
            & %$F_{\left(2, 174\right)}$ = 2.916  
            \\
            Demand& SE & 0.339 & 0.324 & 0.282 & 0.278 & 0.290 & 0.312 & 0.320 & 0.319 & 0.339 & \textbf{\bm{$p =$} 0.034} & %\textbf{\bm{$p=$ 0.023}} 
            & %$p=$ 0.057 
            \\
            %\midrule
            \addlinespace
            Temporal& Mean &2.967& 2.833 & 2.700 & 2.667 & 2.667 & 2.733 & 3.867 & 3.133 & 3.600 & $F_{\left(4, 174\right)}$ = 1.727 & $F_{\left(2, 87\right)}$ = 2.369 & $F_{\left(2, 174\right)}$ = 2.296  \\
            Demand& SE & 0.344 & 0.326 & 0.329 & 0.330 & 0.316 & 0.287 & 0.321 & 0.298 & 0.351  & $p$ = 0.146 & $p$ = 0.100 & $p$ = 0.104   \\
            %\midrule
            \addlinespace
            \multirow{1}{*}{Performance}  & Mean & 5.667 & 5.800 & 5.567 & 5.400 & 5.500 & 5.567 & 5.300 & 5.633 & 4.867 & $F_{\left(4, 174\right)}$ = 1.567 & $F_{\left(2, 87\right)}$ = 1.411 & $F_{\left(2, 174\right)}$ = 2.437  \\
            & SE & 0.312 & 0.285 & 0.290 & 0.313 & 0.283 & 0.243 & 0.210 & 0.237 & 0.291 & $p$ = 0.185 & $p$ = 0.472 & $p$ = 0.090 \\
            %\midrule
            \addlinespace
            \multirow{1}{*}{Effort} & Mean & 3.100 & 3.300 & 3.033 & 3.200 & 3.133 & 3.433 & 4.733 & 4.033 & 4.433 & $F_{\left(4, 174\right)}$ = 2.540 & %$F_{\left(2, 87)}$ = 5.930 
            & %$F_{\left(2, 174)}$ = 1.076  
            \\
            & SE & 0.305 & 0.322 & 0.327 & 0.344 & 0.335 & 0.341 & 0.275 & 0.309 & 0.274  & \textbf{\bm{$p =$} 0.042} & %\textbf{\bm{$p=$ 0.004}} 
            &%$p=$ 0.343 
            \\
            \addlinespace
            \multirow{1}{*}{Frustration} & Mean & 2.467 & 2.233 & 2.467 & 2.367 & 2.367 & 2.500 & 3.433 & 2.567 & 3.067 & $F_{\left(4, 174\right)}$ = 1.847 & $F_{\left(2, 87\right)}$ = 1.829 & $F_{\left(2, 174\right)}$ = 4.082  \\
            & SE & 0.302 & 0.278 & 0.287 & 0.301 & 0.320 & 0.299 & 0.317 & 0.257 & 0.318 & $p$ = 0.122 & $p$ = 0.169 & \textbf{\bm{$p=$} 0.019}  \\
            \bottomrule
        \end{tabular}
            \\
            \\
            \begin{tabular}{clrrr}
            \multicolumn{5}{c}{Post hoc Analysis Results} \\
            \toprule
                ~ & ~ & Narrow-Medium & Narrow-Wide & Medium-Wide \\
                \midrule
                Frustration & Mean Difference & 0.367 & -0.171 & -0.207\\
                 & SE & 0.128 & 0.138 & 0.139\\
                & $p$ & \textbf{0.016} & 1.000  & 0.124\\
                \bottomrule
            \end{tabular}
    \end{tabular}
    }
    %\end{adjustbox}
\end{table}

\begin{table}[ht]
    \centering
    \resizebox{\textwidth}{!}{
            \begin{tabular}{clrrrrrrrrr}
                \multicolumn{11}{c}{Pairwise comparisons for each of the three Node Quantity}\\
                \toprule
                & &  \multicolumn{3}{c}{Narrow Node Width} & \multicolumn{3}{c}{Medium Node Width} & \multicolumn{3}{c}{Wide Node Width} \\
                
                \cmidrule(lr){3-5} \cmidrule(lr){6-8} \cmidrule(lr){9-11}
                & & \multicolumn{1}{c}{Low-Medium} & \multicolumn{1}{c}{Low-High} & \multicolumn{1}{c}{Medium-High} & \multicolumn{1}{c}{Low-Medium} & \multicolumn{1}{c}{Low-High} & \multicolumn{1}{c}{Medium-High} & \multicolumn{1}{c}{Low-Medium} & \multicolumn{1}{c}{Low-High} & \multicolumn{1}{c}{Medium-High}\\
                \cmidrule(lr){1-11}
                Mental & Mean Difference & 0.100 & -1.333 & -1.433 & 0.067 & -0.667 & -0.733 & 0 & -1.333 & -1.333
                \\
                Demand & SE & 0.493 & 0.493 & 0.493 & 0.491 & 0.491 & 0.491 & 0.469 & 0.469 & 0.469 
                \\
                & $p$ & 1.000 & \textbf{0.025} & \textbf{0.014} & 1.000 & 0.533 & 0.416 & 1.000 & \textbf{0.017} & \textbf{0.017}
                \\
                %\midrule
                \addlinespace
                Physical& Mean Difference & 0.367 & -0.933 & -1.300 & 0.167 & -0.700 & -0.867 & -0.367 & -1.300 & -0.933
                \\
                Demand& SE & 0.444 & 0.444 & 0.444 & 0.440 & 0.440 & 0.440 & 0.441 & 0.441 & 0.441 
                \\ 
                & $p$ &  1.000 & 0.115 & \textbf{0.013} & 1.000 & 0.346 & 0.156 & 1.000 & \textbf{0.012} & 0.112
                \\
                \addlinespace
                \multirow{1}{*}{Effort} & Mean Difference & -0.100 & -1.633 & -1.533 & 0.167 & -0.733 & -0.900 & -0.400 & -1.400 & -1.000
                \\
                & SE & 0.437 & 0.437 & 0.437 & 0.455 & 0.455 & 0.455 & 0.446 & 0.446 & 0.446
                \\
                & $p$ &  1.000 & \textbf{\bm{$<$} 0.001} & \textbf{0.002} & 1.000 & 0.333 & 0.154 & 1.000 & \textbf{0.007} & 0.083
                \\
                \bottomrule
                \end{tabular}
        
    \label{tab:my_label}
    }
\end{table}

\begin{table}[ht]
    \centering
    \resizebox{\textwidth}{!}{
            \begin{tabular}{clrrrrrrrrr}
                \multicolumn{11}{c}{Pairwise comparisons for each of the three Node Width}\\
                \toprule
                & & \multicolumn{3}{c}{Low Node Quantity} & \multicolumn{3}{c}{Medium Node Quantity} & \multicolumn{3}{c}{High Node Quantity} \\
                \cmidrule(lr){3-5} \cmidrule(lr){6-8} \cmidrule(lr){9-11}
                & &  \multicolumn{1}{c}{Narrow-Medium} & \multicolumn{1}{c}{Narrow-Wide} & \multicolumn{1}{c}{Medium-Wide} & \multicolumn{1}{c}{Narrow-Medium} & \multicolumn{1}{c}{Narrow-Wide} & \multicolumn{1}{c}{Medium-Wide} & \multicolumn{1}{c}{Narrow-Medium} & \multicolumn{1}{c}{Narrow-Wide} & \multicolumn{1}{c}{Medium-Wide}  \\
                \cmidrule(lr){1-11}
                Mental & Mean Difference & 0.133 & 0.133 & 0 & 0.100 & 0.033 & -0.067 & 0.800 & 0.133 & -0.667
                \\
                Demand & SE & 0.199 & 0.184 & 0.204 & 0.199 & 0.184 & 0.204 & 0.199 & 0.184 & 0.204
                \\
                & $p$ &  1.000 & 1.000 & 1.000 & 1.000 & 1.000 & 1.000 & \textbf{\bm{$<$} 0.001} & 1.000 & \textbf{0.005}
                \\
                %\midrule
                \addlinespace
                Physical& Mean Difference & 0.233 & 0.400 & 0.167 & 0.033 & -0.333 & -0.367 & 0.467 & 0.033 & -0.433
                \\
                Demand& SE & 0.207 & 0.189 & 0.173 & 0.207 & 0.189 & 0.173 & 0.207 & 0.189 & 0.173
                \\
                & $p$ & 0.788 & 0.110 & 1.000 & 1.000 & 0.242 & 0.112 & 0.080 & 1.000 & \textbf{0.043}
                \\
                \addlinespace
                \multirow{1}{*}{Effort} & Mean Difference & -0.200 & 0.067 & 0.267 & 0.067 & -0.233 & -0.300 & 0.700 & 0.300 & -0.400 
                \\
                & SE & 0.241 & 0.230 & 0.229 & 0.241 & 0.230 & 0.229 & 0.241 & 0.230 & 0.229
                \\
                & $p$ & 1.000 & 1.000 & 0.745 & 1.000 & 0.937 & 0.583 & \textbf{0.014} & 0.584 & 0.254
                \\
                \bottomrule
                \end{tabular}
        }
\end{table}

\newpage
\section{Summary of Statistical Results from Phase III}
\label{sec:app:phaseIII}

\subsection{Completion Time for Different Design Choices}
\begin{table}[ht]
    \centering
    % \caption{Experiment II Completion Time(sec)}
    \resizebox{0.7\textwidth}{!}{
        \begin{tabular}{clrrrrr}
            \toprule
            & & \multicolumn{1}{c}{Baseline} & \multicolumn{1}{c}{Tick Marks} & \multicolumn{1}{c}{Color Gradient} & \multicolumn{1}{c}{Combined} & \multicolumn{1}{c}{Friedman Test}   \\
            \cmidrule(lr){3-6}  \cmidrule(lr){7-7} 
            Existence & Mean & 15.626  & 16.305  & 15.016  & 17.461 & \(\chi^2\left(3\right) = 2.000 \)  \\
            Verification & SE & 1.319  & 1.044  & 0.696  & 1.294  & $p$ = 0.572    \\
            & Mean Rank & 2.250  & 2.580  & 2.420  & 2.750 &   \\
            \addlinespace
            Criteria & Mean & 15.757  & 21.004  & 17.727  & 17.239  & \(\chi^2\left(3\right) = 6.100 \)    \\ 
            Matching & SE & 1.367  & 1.844  & 0.876  & 0.923  & $p$ = 0.107   \\
            & Mean Rank & 2.040  & 2.960  & 2.540  & 2.460 &   \\
            \addlinespace
            Comparative & Mean & 28.022  & 26.355  & 28.572  & 30.156 & \(\chi^2\left(3\right) = 4.350 \)    \\
            Analysis & SE & 2.578  & 1.916  & 1.283  & 2.134  & $p$ = 0.226    \\
             & Mean Rank & 2.290  & 2.170  & 2.790  & 2.750 &   \\
            \addlinespace
            Extremes & Mean & 10.487  & 10.716  & 10.851  & 11.898  & \(\chi^2\left(3\right) = 4.650 \)    \\
            Identification & SE & 0.886  & 0.705  & 0.562  & 0.647  & $p$ = 0.199    \\
             & Mean Rank & 2.040  & 2.670  & 2.500  & 2.790 &   \\
            \addlinespace
            Connection & Mean & 11.633  & 11.890  & 14.137  & 10.826  & \(\chi^2\left(3\right) = 0.750 \)    \\ 
            Counting & SE & 0.886  & 0.881  & 2.578  & 0.646 & $p$ = 0.861    \\
             & Mean Rank & 2.330  & 2.580  & 2.630  & 2.460 &   \\
            \bottomrule
        \end{tabular}
    }
\end{table}

\subsection{Error Occurrences for Different Design Choices}
\begin{table}[ht]
    \centering
    % \caption{Experiment II Error Occurrences}
    \resizebox{0.7\textwidth}{!}{
        \begin{tabular}{clrrrrr}
            \toprule
            & & \multicolumn{1}{c}{Baseline} & \multicolumn{1}{c}{Tick Marks} & \multicolumn{1}{c}{Color Gradient} & \multicolumn{1}{c}{Combined} & \multicolumn{1}{c}{Friedman Test}   \\
            \cmidrule(lr){3-6}  \cmidrule(lr){7-7} 
            Average & Mean & 0.017  & 0.017  & 0.050  & 0.017    & \(\chi^2\left(3\right) = 3.375 \)  \\
            Error & SE & 0.012  & 0.012  & 0.025  & 0.012  & $p$ = 0.337   \\
            & Mean Rank & 2.440  & 2.440  & 2.690  & 2.440 &   \\
            \bottomrule
        \end{tabular}
    }
\end{table}

\subsection{Subjective Ratings on Workload}
\begin{table}[ht]
    \centering
    % \caption{Experiment II Workload}
    \resizebox{0.7\textwidth}{!}{
        \begin{tabular}{clrrrrr}
            \toprule
            & & \multicolumn{1}{c}{Baseline} & \multicolumn{1}{c}{Tick Marks} & \multicolumn{1}{c}{Color Gradient} & \multicolumn{1}{c}{Combined} & \multicolumn{1}{c}{Friedman Test}   \\
            \cmidrule(lr){3-6}  \cmidrule(lr){7-7} 
            Mental & Mean & 2.875  & 3.000  & 3.167  & 3.042  & \(\chi^2\left(3\right) = 3.669 \)  \\
            Demand & SE & 0.337  & 0.351  & 0.322  & 0.332  & $p$ = 0.300   \\
            & Mean Rank & 2.270  & 2.380  & 2.790  & 2.560  &   \\
            \addlinespace
            Physical & Mean & 2.250  & 2.417  & 2.333  & 2.417  & \(\chi^2\left(3\right) = 2.012 \)    \\ 
            Demand & SE & 0.250  & 0.262  & 0.238  & 0.269  & $p$ = 0.570   \\
            & Mean Rank & 2.330  & 2.500  & 2.520  & 2.650  &   \\
            \addlinespace
            Temporal & Mean & 2.500  & 2.500  & 2.500  & 2.625  & \(\chi^2\left(3\right) = 0.833 \)    \\
            Demand & SE & 0.313  & 0.295  & 0.301  & 0.275  & $p$ = 0.841   \\
             & Mean Rank & 2.480  & 2.460  & 2.420  & 2.650  &   \\
            \addlinespace
            \multirow{1}{*}{Performance} & Mean & 5.708  & 5.750  & 5.583  & 5.708  & \(\chi^2\left(3\right) = 1.903 \)    \\
            & SE & 0.259  & 0.250  & 0.240  & 0.221  & $p$ = 0.593   \\
             & Mean Rank & 2.520  & 2.670  & 2.350  & 2.460  &   \\
            \addlinespace
            \multirow{1}{*}{Effort} & Mean & 3.583  & 3.542  & 3.500  & 3.625  & \(\chi^2\left(3\right) = 0.195 \)    \\ 
            & SE & 0.366  & 0.340  & 0.366  & 0.360  & $p$ = 0.978   \\
             & Mean Rank & 2.480  & 2.440  & 2.520  & 2.560  &   \\
            \addlinespace
            \multirow{1}{*}{Frustration} & Mean & 2.042  & 2.042  & 2.250  & 2.250  & \(\chi^2\left(3\right) = 3.988\)   \\
            & SE & 0.259  & 0.252  & 0.290  & 0.250  & $p$ = 0.263   \\
             & Mean Rank & 2.330  & 2.350  & 2.650  & 2.670  &   \\
            \bottomrule
        \end{tabular}
    }
\end{table}

\end{document}